\documentclass[aps,preprintnumbers,superscriptaddress,showpacs]{revtex4}
\usepackage{epsfig}
\usepackage{psfrag}
\usepackage{amsfonts}
\usepackage{graphicx}
\usepackage{dcolumn}
\usepackage{bm}
\usepackage{amssymb}
\usepackage{amsmath}
\usepackage{float}
\usepackage{xcolor}
\begin{document}

\title{Heavy quarkonium spectral function in an anisotropic background}

\author{Wen-Bin Chang}
\email{changwb@mails.ccnu.edu.cn} \affiliation{Institute of Particle Physics and Key Laboratory of Quark and Lepton Physics (MOS), Central China Normal University,
Wuhan 430079,China}

\author{De-fu Hou}
\email{houdf@mail.ccnu.edu.cn} \affiliation{Institute of Particle Physics and Key Laboratory of Quark and Lepton Physics (MOS), Central China Normal University,
Wuhan 430079,China}

\begin{abstract}

 In this paper, we use a five-dimensional Einstein-dilaton-two-Maxwell holographic QCD model to investigate the dissociation effects of $J/\Psi$ and $\Upsilon(1S)$ states in an anisotropic medium by calculating their spectral functions.
First, we present the holographic quarkonium masses at zero temperature via Physics-Informed Neural Networks.
Then, at finite temperature, we derive the spectral functions, representing heavy vector mesons as peaks, and observe that with increasing anisotropy, temperature, chemical potential, and warp factor, the peak height diminishes while its width expands, indicating an accelerated dissociation process.
Additionally, the results indicate the anisotropy induces a stronger dissociation effect in the direction parallel to the polarization compared to the perpendicular, revealing the anisotropy’s directional influence.

\end{abstract}

\maketitle

\section{Introduction}
Heavy quarkonium is the bound state of a heavy quark and its antiquark, such as $J/\Psi$ and $\Upsilon(1S)$, which are composed of charm and bottom quarks, respectively.
In heavy ion collisions, heavy flavor quarks, either newly produced or excited from the vacuum, play an irreplaceable role in the evolution of the Quark-Gluon Plasma (QGP), interacting strongly with the QGP medium constituents or even becoming part of the QGP itself, positioning heavy flavor hadrons as unique probes into the properties of the QGP and the nature of strong interaction \cite{jr,hva,hva2,amo,mhe,pbr,nxu,zl}.
Historically, the notion of quarkonium suppression can be traced back to the work of Ref. \cite{tma}, where the bound state created in the early stages of the heavy ion collision eventually disintegrates into unbound quarks when traversing the QGP medium, leading to a decrease in the dilepton spectrum due to the color screening.
Thus, investigating heavy quarkonium mesons, which undergo dissociation in the QGP medium that influences their final yields and spectra, is crucial for comprehending the formation and evolution of QGP.

The experiments have revealed that the QGP exhibits the properties of a strongly coupled fluid, challenging the applicability of perturbative methods and necessitating non-perturbative approaches \cite{esh,esh2}.
One of the theoretical frameworks is gauge/gravity duality, also known as holography, which relates a strongly coupled gauge theory to a weakly coupled gravity theory in a higher dimensional space \cite{mald,witt,irke}.
Utilizing this approach, one can employ classical gravity solutions to compute non-perturbative quantities in gauge theory, which are traditionally challenging using perturbative methods.
Various holographic models have been successfully applied to study heavy flavor physics in QGP.
In Ref. \cite{pco,asm}, scalar glueballs and scalar mesons at finite temperature are examined within the holographic QCD framework.
In studies \cite{ddu,lah,yq}, the soft-wall model is employed to examine the effect of a background magnetic field on vector meson melting, while the spectrum of vector mesons in finite temperature plasma and their dependence on temperature and momentum are analyzed, respectively.
Using a consistent AdS/QCD approach, the masses and decay constants of charmonium and bottomonium states are described in works \cite{nrf,nrf2}.
Refs. \cite{nrf3,nrf4,nrf5,lah2} present a holographic bottom up model for the thermal behavior of heavy vector mesons inside a plasma at finite temperature and density, while the meson dissociation in the medium is represented by the decrease in the height of the spectral function peaks.
The quasinormal modes for charmonium and bottomonium, which are gravity solutions representing the quasiparticle states in the thermal medium, have been studied, and a consistent description of the dissociation process is found in \cite{nrf6,nrf7,yq2}.
In Ref. \cite{yq3}, the authors investigate the impact of temperature, chemical potential, and rotation on the dissociation of $J/\Psi$ in magnetized, rotating QGP matter, highlighting an augmented dissociation effect and increased effective mass in the QGP phase.
Further illuminating works are detailed in references \cite{yk,xc,sli,rzo}.

However, most of the existing studies of heavy quarkonium have assumed that the QGP is isotropic and homogeneous, i.e., that it has no preferred direction, and this is not necessarily true in realistic situations.
Experimental data evidence from heavy ion collisions suggests an early stage local anisotropy in QGP, with the system predominantly expanding along the collision axis \cite{mst,dgi}.
To understand how the anisotropy of QGP modifies its properties and observables, in this paper we investigate the spectral functions of heavy quarkonium mesons in an anisotropic background.
A bottom-up approach is used to consider the five-dimensional gravity dual of a strongly coupled anisotropic QGP with finite temperature and chemical potential, where a spatial anisotropy in the $x_1$ direction is specified by the parameter $\nu$ that reduces to the isotropic case when $\nu = 1$.
Our paper has the following structure.
In section II, we review the anisotropic black brane solutions obtained in \cite{aref}, supported by the Einstein-dilaton-two-Maxwell action.
The detailed calculation of the spectral functions is presented in Section III.
The final section presents our results and discussions.

\section{Anisotropic Holographic Model}
In the following, we briefly review the anisotropic holographic model introduced in \cite{aref}, which is based on the 5-dimensional Einstein-dilaton-two-Maxwell system. The Einstein frame action of the system, denoted as $S$, is formulated as 
\begin{equation}
S=\int \frac{d^5 x}{16 \pi G_5} \sqrt{-\operatorname{det}\left(g_{\mu \nu}\right)}\left[R-\frac{f_1(\phi)}{4} F_{(1)}^2-\frac{f_2(\phi)}{4} F_{(2)}^2-\frac{1}{2} \partial_\mu \phi \partial^\mu \phi-V(\phi)\right].\label{sng}
\end{equation}
The chemical potential and anisotropy are introduced by two Maxwell fields $F_{(1)}$ and $F_{(2)}$, with field strength tensors
$F_{\mu \nu}^{(1)}=\partial_\mu A_\nu-\partial_\nu A_\mu$ and 
$F_{\mu \nu}^{(2)}=q d y^1 \wedge d y^2$, respectively.
The dilaton field $\phi$, with potential $V(\phi)$, couples to these two Maxwell fields through the gauge kinetic functions $f_1(\phi)$ and $f_2(\phi)$.
To describe the anisotropic background holographically, we adopt the black brane metric ansatz in the following form \cite{aref}
\begin{gather}
  ds^{2} =\frac{L^{2} \, b(z)}{z^{2}}
  \left[ - \ g(z) dt^{2} + dx_1^2 + z^{2-\frac{2}{\nu}} \left(
      dx_{2}^{2} + dx_{3}^{2} \right) + \frac{dz^{2}}{g(z)}
  \right], \label{eq:2.02} \\
  \phi = \phi (z), \qquad
  A_{\mu }^{(1)} = A_{t} (z) \delta _{\mu}^{0}. \qquad\label{eq:2.03}
\end{gather}
Here, $b(z)=e^{c z^2 / 2}$ represents the AdS deformation factor, with $c$ signifying the deviation from conformality.
$g(z)$ is the blackening function, and $L$ is the  characteristic length scale (set to one for convenience).
A non-zero time component, $A_t(0) = \mu$, of the first Maxwell field $F_{(1)}$, serves as the chemical potential in the dual system.
By solving the equation of motion derived from the aforementioned action, the functioncan $g(z)$ be determined as
\begin{equation}
g(z)=1-\frac{z^{2+\frac{2}{\nu}}}{z_{h}^{2+\frac{2}{\nu}}} \frac{\mathfrak{G}\left(\frac{3}{4} c z^{2}\right)}{\mathfrak{G}\left(\frac{3}{4} c z_{h}^{2}\right)}-\frac{\mu^{2} c z^{2+\frac{2}{\nu}} e^{\frac{c z_{h}^{2}}{2}}}{4\left(1-e^{\frac{c z_{h}^{2}}{4}}\right)^{2}} \mathfrak{G}\left(c z^{2}\right)+\frac{\mu^{2} c z^{2+\frac{2}{\nu}} e^{\frac{c z_{h}^{2}}{2}}}{4\left(1-e^{\frac{c z_{h}^{2}}{4}}\right)^{2}} \frac{\mathfrak{G}\left(\frac{3}{4} c z^{2}\right)}{\mathfrak{G}\left(\frac{3}{4} c z_{h}^{2}\right)} \mathfrak{G}\left(c z_{h}^{2}\right),
\end{equation}
where
\begin{equation}
\mathfrak{G}(x)=\sum_{n=0}^{\infty} \frac{(-1)^{n} x^{n}}{n !\left(1+n+\frac{1}{\nu}\right)}.
\end{equation}

In this anisotropic 5-dimensional solution, the anisotropy parameter $\nu$ serves as an arbitrary dynamical exponent. However, in accordance with the experimental multiplicity data provided in reference \cite{ALICE}, $\nu$ is confined within the range of $1$ to $4.5$ \cite{aref,shock}.

Consequently, the temperature can be obtained as follows: 
\begin{equation}
T\left(z_{h}, \mu, c, \nu\right)=\frac{g^{\prime}\left(z_{h}\right)}{4 \pi}=\frac{e^{-\frac{3 c z_{h}^{2}}{4}}}{2 \pi z_{h}}\left|\frac{1}{\mathfrak{G}\left(\frac{3}{4} c z_{h}^{2}\right)}+\frac{\mu^{2} c z_{h}^{2+\frac{2}{\nu}} e^{\frac{c z_{h}}{4}}}{4\left(1-e^{\frac{c z_{h}^{2}}{4}}\right)^{2}}\left(1-e^{\frac{c z_{h}^{2}}{4}} \frac{\mathfrak{G}\left(c z_{h}^{2}\right)}{\mathfrak{G}\left(\frac{3}{4} c z_{h}^{2}\right)}\right)\right|.
\end{equation}
For chemical potentials below the critical value, $\mu<\mu_{cr}$, the temperature function $T(z_h)$ is multivalued, while for $\mu \geq \mu_{cr}$, it is single-valued (see \cite{aref} for the detailed thermodynamics of this holographic model).

\section{SPECTRAL FUNCTION}
In this study, we investigate the spectral functions for heavy quarkonium, which are calculated using the phenomenological model \cite{nrf4} that represents heavy quarkonium by a vector field 
$V_m=(V_\mu, V_z)(\mu=0,1,2,3)$ dual to the electric current operator 
$J^\mu=\overline{\Psi}\gamma^\mu\Psi$.
The calculation is carried out using the membrane paradigm \cite{niq}, with the bulk action for the vector field described by 
\begin{equation}
S=-\int d^5 x \sqrt{-g} \frac{e^{-\phi(z)}}{4 g_5^2} F^{m n} F_{m n},
\end{equation} where $F_{mn}=\partial_mV_n-\partial_nV_m$.
This phenomenological model includes parameters $\kappa$, $\Gamma$, $M$ and $\alpha$ for quark mass, string tension, non-hadronic decay, and  constituent quark mass effect to describe the heavy quarkonium states  within the dilaton field \cite{nrf8}
\begin{equation}\label{eq07}
  \phi(z)=(\kappa z)^{2-\alpha}+Mz+\tanh(\frac{1}{Mz}-\frac{\kappa}{\sqrt{\Gamma}}).
\end{equation}
The values of these parameters for charmonium and bottomonium are respectively :
\begin{align}\label{eq08}
  \kappa_c & =1.8GeV,\quad \sqrt{\Gamma_c}=0.53GeV,\quad M_c=1.7GeV,\quad \alpha_c=0.54,\notag \\
  \kappa_b & =9.9GeV,\quad \sqrt{\Gamma_b}=1.92GeV,\quad M_b=2.74GeV,\quad \alpha_b=0.863.
\end{align}
The equation of motion obtained from the bulk action is
\begin{equation}\label{eom}
  \partial^{m}(\frac{\sqrt{-g}}{e^{\phi(z)}}F_{mn})=0,
\end{equation}
and the conjugate momentum of the gauge field with respect to the $z$-foliation is given by
\begin{equation}\label{eq11}
 j^\mu=-\frac{\sqrt{-g}F^{z\mu}}{e^{\phi(z)}}.
\end{equation}
Assuming a plane wave resolution for the vector field, we propose its propagation in the direction of anisotropy. 
Because the anisotropy is in the $x_1$ axis direction, these solutions should not be influenced by the perpendicular coordinates $x_2$ and $x_3$. 
This allows us to separate the equations of motion \eqref{eom} into two distinct channels: one longitudinal channel depicting fluctuations  along $(t, x_1)$ and a transverse one showing fluctuations along the spatial directions $(x_2, x_3)$, which will yield different flow equations for the vector field components.
The relevant longitudinal components $t, x_1$ and $z$ of Eq. \eqref{eom} in this case read
\begin{align}
  -\partial_zj^t-\frac{\sqrt{b(z)}e^{-\phi(z)}z^{\frac{-2+\nu}{\nu}}}{g(z)}\partial_{x_1}F_{x_1t} & =0, \label{eq11}\\
  -\partial_zj^{x_1}+\frac{\sqrt{b(z)}e^{-\phi(z)}z^{\frac{-2+\nu}{\nu}}}{g(z)}\partial_tF_{x_1t} & =0,\label{eq12} \\
  \partial_{x_1}j^{x_1}+\partial_tj^t & =0\label{eq13}.
\end{align}
Applying the Bianchi identity enables us to achieve
\begin{equation}\label{eq14}
  \partial_zF_{x_1t}-\frac{e^{\phi(z)}z^{-1+\frac{2}{\nu}}}{\sqrt{b(z)}g(z)}\partial_tj^{x_1}-\frac{e^{\phi(z)}z^{-1+\frac{2}{\nu}}}{\sqrt{b(z)}}\partial_{x_1}j^t=0.
\end{equation}
We designate the longitudinal conductivity  in terms of
\begin{align}
  \sigma_L(\omega,\overrightarrow{p},z) & =\frac{j^{x_1}(\omega,\overrightarrow{p},z)}{F_{x_1t}(\omega,\overrightarrow{p},z)},\label{eq15} \\
  \partial_z \sigma_L(\omega,\overrightarrow{p},z) & =\frac{\partial_zj^{x_1}}{F_{x_1t}}-\frac{j^{x_1}}{F_{x_1t}^2}\partial_zF_{x_1t}\label{eq16}.
\end{align}
Employing Kubo formula, we establish a connection between the five-dimensional conductivity at $z=0$ and the retarded Green's function:
\begin{equation}
  \sigma_L(\omega)=\frac{-G_R^L(\omega)}{i\omega}.
\end{equation}
For a plane wave solution with momentum $P=(\omega,0,0,0)$, we use  Eq.\eqref{eq12}, Eq.\eqref{eq13}, and Eq.\eqref{eq14} to derive the flow equation for the longitudinal channel
\begin{equation}
 \partial_z \sigma_L=\frac{i\omega}{g(z)\sqrt{b(z)}e^{-\phi(z)}z^{1-\frac{2}{\nu}}}(\sigma_L^2-b(z)e^{-2\phi(z)}z^{2-\frac{4}{\nu}}).\label{eq18}
\end{equation}
Using a similar process, we can obtain the flow equation for the transverse channel:
\begin{equation}
 \partial_z \sigma_T=\frac{iz\omega}{g(z)\sqrt{b(z)}e^{-\phi(z)}}(\sigma_T^2-\frac{b(z)e^{-2\phi(z)}}{z^2}).\label{eq19}
\end{equation}
The regularity condition at the horizon is employed to solve the equations with $\partial_z \sigma_{L(T)}=0$, and the spectral function can then be obtained from the imaginary part of the retarded Green’s functions
\begin{equation}\label{eq20}
  \rho(\omega)\equiv-Im G_R(\omega)=\omega Re\,\sigma(\omega,0).
\end{equation}
Notably, under the condition that the anisotropy parameter $\nu$ was set to 1, the metric became isotropic, resulting in both flow equations \eqref{eq18} and \eqref{eq19} having the same form.

\section{RESULTS AND DISCUSSION}
Using the dilaton model with \eqref{eq07}, we start by investigating the holographic quarkonium masses of the charmonium and bottomonium at zero temperature, following the phenomenological approach presented in reference \cite{nrf4}.
Adopting the gauge $V_z=0$, the equation for the transverse components has normalizable solutions in momentum space, given by
\begin{equation}
\partial_z\left(e^{-B} \partial_z V_n\right)+m_n^2 e^{-B} V_n=0,\label{eq20}
\end{equation} where $B=\phi(z)-\frac{1}{2} \log(\frac{\mathrm{b}(z)}{z^2})$ and $m_n$ are the meson masses.
We can rewrite equation \eqref{eq20} as a Schrödinger-like equation
\begin{equation}
\begin{gathered}
-\psi_n^{\prime \prime}+U(z) \psi_n=m_n^2 \psi_n, \\
U(z)=\frac{1}{4}\left(B^{\prime}\right)^2-\frac{1}{2} B^{\prime \prime},\label{eq23}
\end{gathered}
\end{equation}
with the Bogoliubov transformation $V_n=e^{B/2} \psi_n$, and solving this equation results in the quarkonium masses \cite{aka}.

Physics-Informed Neural Networks (PINNs) is an innovative approach that embeds physical laws directly into neural network training by treating these laws as regularization terms or constraints within the loss function.
It presents a powerful methodology across diverse domains, including fluid dynamics and quantum mechanics, offering some advantages over conventional numerical methods, such as mitigating the curse of dimensionality, minimizing numerical error accumulation, and processing differential equations without specific requirements, provided their solutions are finite and regular \cite{pgr,mra,chenx,Teuko}.
Jin, Mattheakis, and Protopapas have recently developed a PINNs algorithm that simultaneously learns eigenvalues and eigenfunctions through a scanning mechanism, effectively solving the quantum eigenvalue problem for finite wells, harmonic oscillators, and hydrogen atom systems by progressively identifying multiple eigenvalues in a single training session \cite{jin1,jin2}.

In this study, we present an extension of the JMP algorithm to tackle more complex potential function scenarios, generalizing its application to solve the mass of quarkonium.
Further details on the JMP algorithm used in our calculations are provided in Appendix \ref{appendixA}.
Figures 1 and 2 depict the evolution of the loss function and show the successful recognition of the four lowest eigenvalues of  charmonium and bottomonium by our neural network.
\begin{figure}[H]
\centering
\includegraphics[width=8cm]{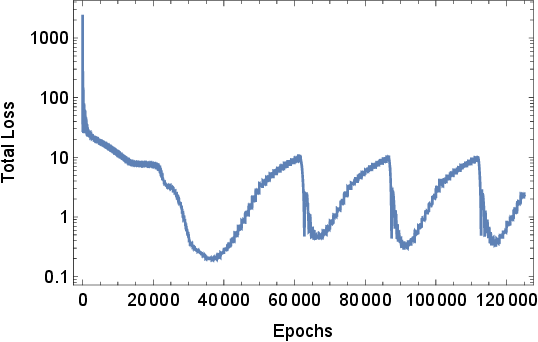}
\includegraphics[width=8cm]{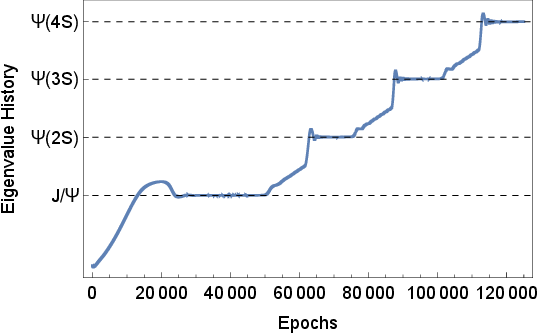}
\caption{The network’s training process for charmonium.
Left:  The total loss function versus training epochs. Right: The predicted eigenvalues of the network are represented by dotted horizontal lines over training epochs.
}
\end{figure}
\begin{figure}[H]
\centering
\includegraphics[width=8cm]{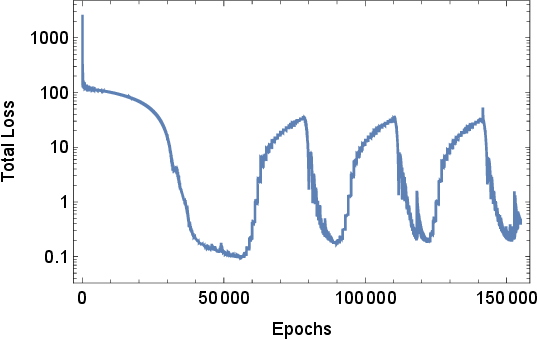}
\includegraphics[width=8cm]{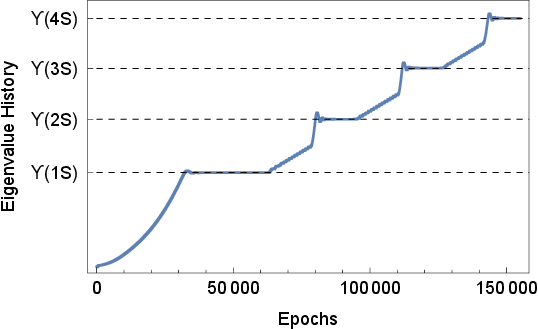}
\caption{The network’s training process for bottomonium.
Left:  The total loss function versus training epochs. Right: The predicted eigenvalues of the network are represented by dotted horizontal lines over training epochs.
}
\end{figure}
The neural network identifies eigenvalues by searching for plateaus, which correspond to distinct depressions in the loss function, with a jump caused by changing $c$ to encourage finding the next eigenvalue.
The summarized results of the neural network for charmonium and bottomonium are shown in tables \ref{table1} and \ref{table2}, respectively, with the experimental data read from PDG \cite{pdg} for comparison.
\begin{table}[ht]
\centering
\begin{tabular}{l |c|c}
\hline 
\hline
 State & Holographic model (MeV) & Charmonium  experimental (MeV)\\
\hline 
 $J/\Psi$ & 2526.49   & $3096.916\pm 0.011$  \\
 $\Psi(2S)$ & 3730.02  & $3686.109\pm 0.012$  \\
 $\Psi(3S)$ & 4281.93  & $4039\pm 1$ \\
 $\Psi(4S)$ & 4754.54  & $4421\pm 4$  \\
\hline\hline
\end{tabular}
\caption{Holographic masses of the Charmonium S-wave resonances, with experimental values from PDG \cite{pdg} serving as a benchmark.
}
\label{table1}
\end{table}

\begin{table}[ht]
\centering
\begin{tabular}{l |c|c}
\hline 
\hline
 State & Holographic model (MeV) & Bottomonium  experimental (MeV)\\
\hline 
 $\Upsilon(1S)$ & 9611.38  & $9460.3\pm0.26$  \\
 $\Upsilon(2S)$ & 10071.76  & $10023.26\pm0.32$  \\
 $\Upsilon(3S)$ & 10469.23  & $10355.2\pm0.5$ \\
 $\Upsilon(4S)$ & 10800.61  & $10579.4\pm1.2$  \\
\hline\hline
\end{tabular}
\caption{Holographic masses of the Bottomonium S-wave resonances, with experimental values from PDG \cite{pdg} serving as a benchmark.
}
\label{table2}
\end{table}
The spectral functions for heavy vector mesons at finite temperature and with anisotropy are calculated numerically by solving the equations \eqref{eq18} and \eqref{eq19} with the boundary conditions described in previous sections, using the model parameters given in \eqref{eq08}.
In Fig. 3 and Fig. 4, we show the numerical results of spectral functions for the $J/\Psi$ state and $\Upsilon(1S)$ state across various anisotropy parameters, as derived from the model under consideration, showcasing the anisotropy both parallel and perpendicular to the polarization. 
The peak features of the curve in the spectra represent quasiparticle states, the peak position denotes the quasiparticle mass, and its width is inversely proportional to the decay rate, indicating the stability of these emergent quasiparticles.
\begin{figure}[H]
\centering
\includegraphics[width=8cm]{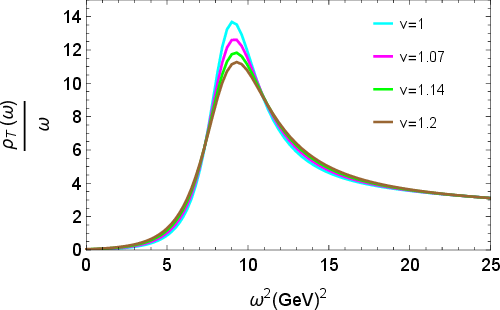}
\includegraphics[width=8cm]{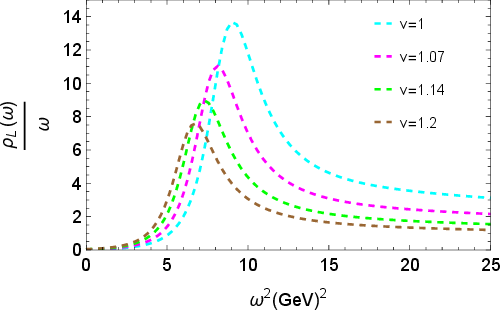}
\caption{Spectral functions for $J/\Psi$ at $T=0.3GeV, \mu=0.1GeV$ and $c=-0.3{GeV}^2$ for different anisotropic parameter.
Left: The anisotropy is perpendicular to the polarization. Right: The anisotropy is parallell to the polarization.
}
\end{figure}
\begin{figure}[H]
\centering
\includegraphics[width=8cm]{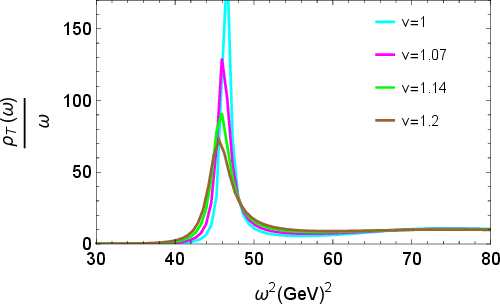}
\includegraphics[width=8cm]{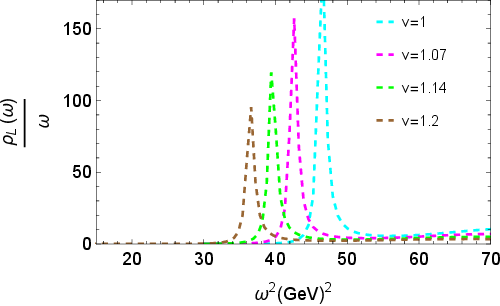}
\caption{Spectral functions for $\Upsilon(1S)$ at $T=0.3GeV, \mu=0.1GeV$ and $c=-0.3{GeV}^2$ for different anisotropic parameter.
Left: The anisotropy is perpendicular to the polarization. Right: The anisotropy is parallell to the polarization.
}
\end{figure}
Our results show that increasing the anisotropy parameter $\nu$ lowers the height and broadens the width of the spectral function peak for both charmonium and bottomonium in parallel and perpendicular scenarios, while also shifting the peak position.
An increased peak width and decreased peak height denote an augmented dissociation effect, suggesting that the presence of anisotropy accelerates the dissociation of heavy quarkonium states, which is in agreement with the calculations of \cite{ugu}.
A physical interpretation of the observation that a larger anisotropy favors quarkonium dissociation is that, at fixed temperature, the screening length between a quark-antiquark pair diminishes with an increase in the anisotropy for any orientation of the dipole, thereby promoting quarkonium dissociation \cite{areb,mch}.
Another intuitive physical picture is that a larger anisotropy can induce larger thermodynamic forces and thus favor quarkonium dissociation.
\begin{figure}[H]
\centering
\includegraphics[width=8cm]{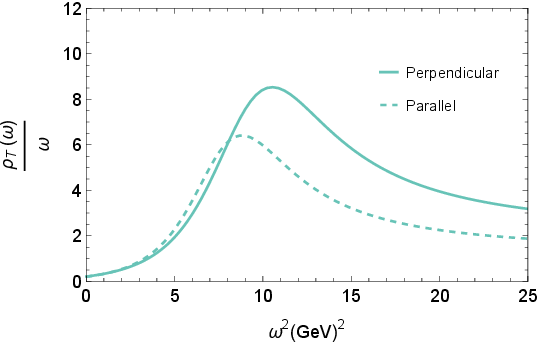}
\includegraphics[width=8cm]{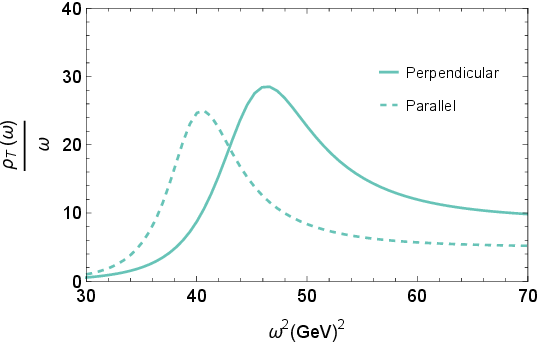}
\caption{Spectral functions for $J/\Psi$ and $\Upsilon(1S)$ with the anisotropy parallel and perpendicular to the polarization at $\mu=0GeV, T=0.4GeV, c=-0.3{GeV}^2$, and $\nu=1.1.$
Left: Spectral functions for $J/\Psi$. Right: Spectral functions for $\Upsilon(1S)$.
}
\end{figure}Comparing the spectral functions of $J/\Psi$ and $\Upsilon(1S)$ states with the anisotropy parallel and perpendicular to the polarization at $\mu=0GeV, T=0.4GeV, c=-0.3{GeV}^2$, and $\nu=1.1$ in Fig. 5, we find that the anisotropy has a stronger dissociation effect when it is parallel to the polarization than when it is perpendicular, since the anisotropy parameter has a greater influence on the anisotropic direction.

In Fig. 6 and Fig. 7, as temperature $T$ rises, the spectral function exhibits a significant reduction in peak height and broadening, whether the anisotropy is parallel or perpendicular to the polarization, indicating an enlarged dissociation effect  with temperature for both $J/\Psi$ and $\Upsilon(1S)$.
\begin{figure}[H]
\centering
\includegraphics[width=8cm]{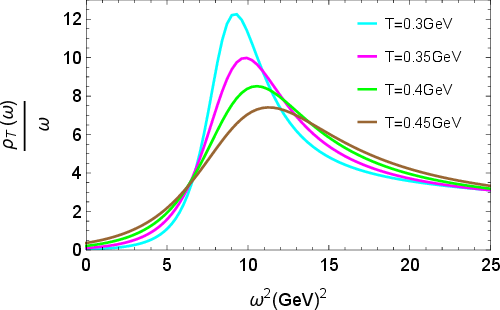}
\includegraphics[width=8cm]{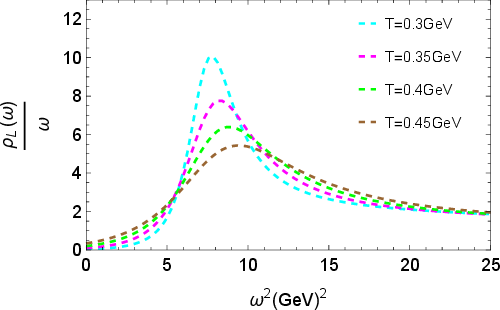}
\caption{Spectral functions for $J/\Psi$ at $\nu=1.1, \mu=0.1GeV$ and $c=-0.3{GeV}^2$ for different temperature.
Left: The anisotropy is perpendicular to the polarization. Right: The anisotropy is parallell to the polarization.
}
\end{figure}
\begin{figure}[H]
\centering
\includegraphics[width=8cm]{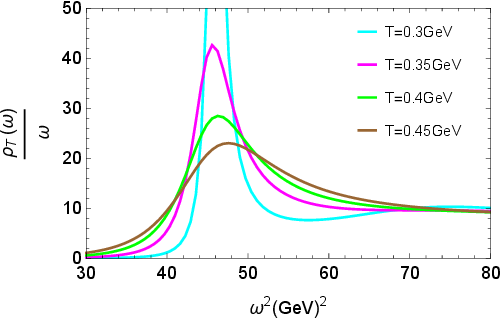}
\includegraphics[width=8cm]{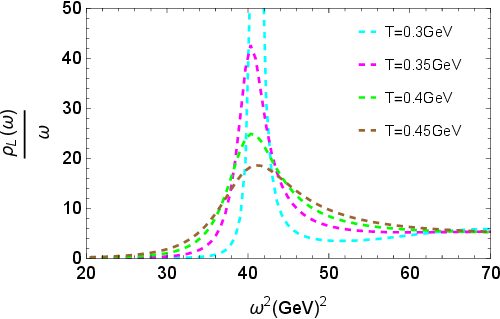}
\caption{Spectral functions for $\Upsilon(1S)$ at $\nu=1.1, \mu=0.1GeV$ and $c=-0.3{GeV}^2$ for different temperature.
Left: The anisotropy is perpendicular to the polarization. Right: The anisotropy is parallell to the polarization.
}
\end{figure}
Similar to the temperature effect, as illustrated in Fig. 8 and Fig. 9, an increase in the chemical potential $\mu$ leads to a decrease in the peak height and a wider width of the spectral function, regardless of the polarization direction, indicating its role in enhancing the dissociation process.
This is consistent with previous studies that have shown that finite temperature and density effects lead to the melting of heavy quarkonium states \cite{nrf3,nrf4,nrf5,lah2}.
\begin{figure}[H]
\centering
\includegraphics[width=8cm]{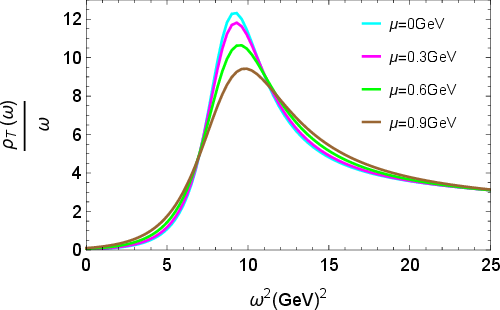}
\includegraphics[width=8cm]{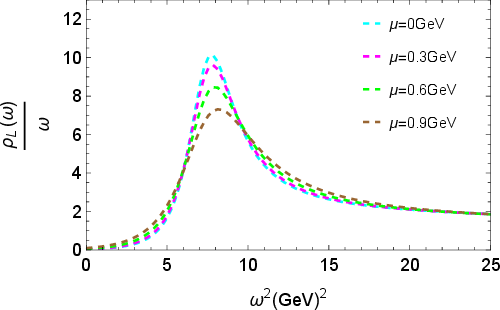}
\caption{Spectral functions for $J/\Psi$ at $T=0.3GeV, \nu=1.1$ and $c=-0.3{GeV}^2$ for different chemical potential.
Left: The anisotropy is perpendicular to the polarization. Right: The anisotropy is parallell to the polarization.
}
\end{figure}
\begin{figure}[H]
\centering
\includegraphics[width=8cm]{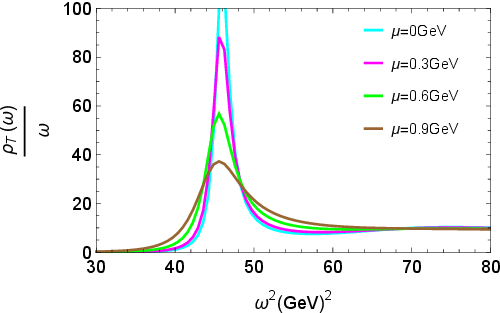}
\includegraphics[width=8cm]{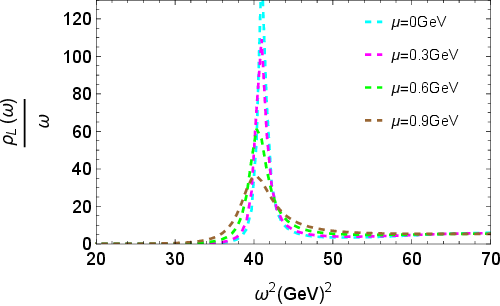}
\caption{Spectral functions for $\Upsilon(1S)$ at $T=0.3GeV, \nu=1.1$ and $c=-0.3{GeV}^2$ for different chemical potential.
Left: The anisotropy is perpendicular to the polarization. Right: The anisotropy is parallell to the polarization.
}
\end{figure}
Further, to elucidate the influence of the warp factor coefficient, we set parameters at $T=0.3GeV, \mu=0.1GeV$ and $\nu=1.1$, with the spectral function variations for distinct $c$ values presented in Fig. 10 and Fig. 11. Elevating the warp factor coefficient diminishes the peak height and expands the spectral function width in both parallel and perpendicular cases, as depicted across various spectral functions in the figures. This suggests that larger warp factor coefficients enhance  the dissociation effect.
\begin{figure}[H]
\centering
\includegraphics[width=8cm]{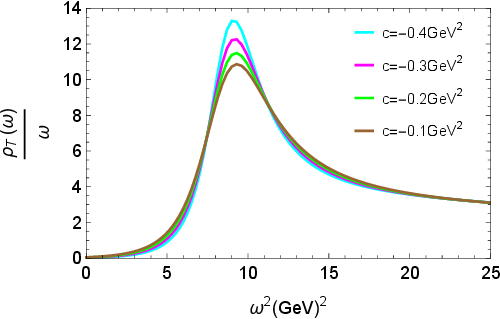}
\includegraphics[width=8cm]{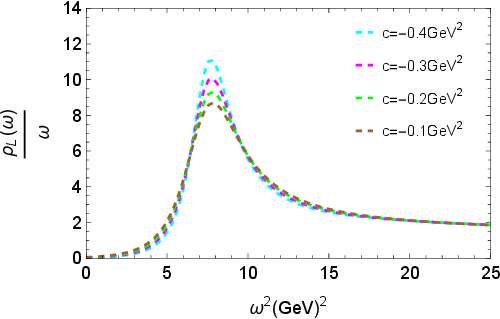}
\caption{Spectral functions for $J/\Psi$ at $T=0.3GeV, \mu=0.1GeV$ and $\nu=1.1$ for different warp factor coefficient.
Left: The anisotropy is perpendicular to the polarization. Right: The anisotropy is parallell to the polarization.
}
\end{figure}
\begin{figure}[H]
\centering
\includegraphics[width=8cm]{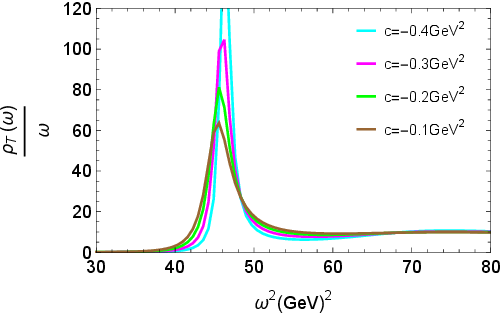}
\includegraphics[width=8cm]{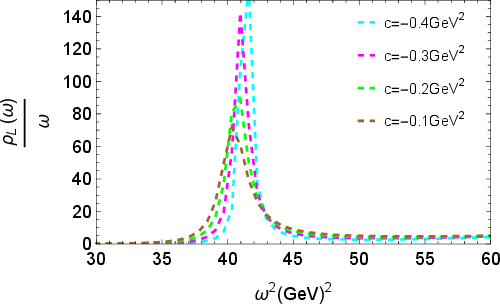}
\caption{Spectral functions for $\Upsilon(1S)$ at $T=0.3GeV, \mu=0.1GeV$ and $\nu=1.1$ for different warp factor coefficient.
Left: The anisotropy is perpendicular to the polarization. Right: The anisotropy is parallell to the polarization.
}
\end{figure}
In this work, we presented the calculation of the spectral function for heavy quarkonium within anisotropic black brane solutions for a bottom-up QCD model. Using numerical calculation, we investigated how the spectral functions of charmonium and bottomonium, revealing the bell shape peaks interpreted as vector mesons in the dual theory, depend on various parameters such as anisotropy, temperature, chemical potential, and warp factor coefficient. Our results reveal that an increased anisotropy accelerates the dissociation of bound states, with a more pronounced effect when the anisotropy is parallel rather than perpendicular to the polarization.
The intuitive physical picture is that:  a larger anisotropy can  induce stronger thermodynamic forces like  shear force for  instance  and thus enhance the quarkonium dissociation.

We also find that as the temperature, chemical potential, and warp factor coefficient rise, the quasiparticle state becomes more unstable, as evidenced by the diminishing peak height and broadening spectral function. Further studies should encompass the influence of rotation on the dissociation effect, as a substantial angular momentum generated by the non-central heavy ion collisions, and we will leave this for further research. 

\section{Acknowledgments} 
We thank Hai-Cang Ren and Yan-Qing Zhao for the fruitful discussions. This work is supported in part by the National Key Research and Development Program of China under Contract No. 2022YFA1604900. This work is also partly supported by the National Natural Science Foundation of China (NSFC) under Grants No. 12275104, No. 11890711, No. 11890710, and No. 11735007.

$$\,$$
\newpage

\appendix
\section*{Appendix}

\section{}\label{appendixA}
Figure 12 illustrates the framework of the JMP algorithm, which solves the eigenvalue problem by learning from its own predictions \cite{jin1,jin2}. 
\begin{figure}[H]
\centering
\includegraphics[width=7cm]{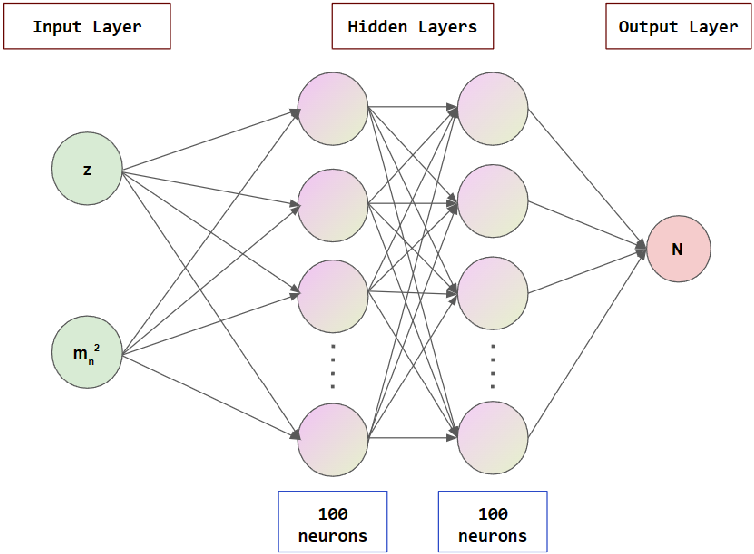}
\caption{The structure of the neural network.
}
\end{figure}
The output of the neural network is denoted by $N(z,m_n^2)$, and the predicted eigenfunction is given by
\begin{equation}
\psi_n(z) = (1 - e^{-(z-z_L)})(1 - e^{-(z-z_R)})N(z,m_n^2), 
\end{equation}
where the left and right boundaries are $z_L$  and $z_R$.
To address equation \eqref{eq23}, the network is optimized through the minimization of a loss function $L$
\begin{equation}
\begin{gathered}
L = \left\langle(-\psi_n^{\prime \prime}(z)+U(z) \psi_n(z)-m_n^2 \psi_n(z)
)^2\right\rangle_z + L_{reg}, \\
L_{reg} = \frac{1}{{\psi_n(z)}^2} +  \frac{1}{(m_n^2)^2} +  e^{-m_n^2+c}.
\end{gathered}
\end{equation}
In our investigation, we used PyTorch to implement a fully-connected neural network with two hidden layers, each comprising $100$ neurons, optimized using the Adam algorithm with a learning rate of $8 \times 10^{-3}$ and sigmoid activation function, with the training set constructed by $2000$ random points within the $[z_L, z_R]$ interval.


\end{document}